# Proposed Next Generation GRB Mission: EXIST

J. Grindlay[1], N. Gehrels[2], F. Harrison[3], R. Blandford[3], G. Fishman[4], C. Kouveliotou[4,] D.H. Hartmann[5], S. Woosley[6], W. Craig[7] and J. Hong[1]

*1. Harvard, 2. NASA/GSFC, 3. Caltech, 4. NASA/MSFC, 5. Clemson Univ., 6. UC Santa Cruz, 7. LLNL*

**Abstract.** A next generation Gamma Ray Burst (GRB) mission to follow the upcoming *Swift* mission is described. The proposed Energetic X-ray Imaging Survey Telescope, *EXIST*, would yield the limiting (practical) GRB trigger sensitivity, broad-band spectral and temporal response, and spatial resolution over a wide field. It would provide high resolution spectra and locations for GRBs detected at GeV energies with *GLAST*. Together with the next generation missions *Constellation-X, NGST* and *LISA* and optical-survey (*LSST*) telescopes, *EXIST* would enable GRBs to be used as probes of the early universe and the first generation of stars. *EXIST* alone would give ~10-50" positions (long or short GRBs), approximate redshifts from lags, and constrain physics of jets, orphan afterglows, neutrinos and SGRs.

## INTRODUCTION

Gamma-ray bursts (GRBs) are the most luminous events since the Big Bang. Current models for at least the "long" GRBs favor their origin in the collapse of massive stars. As such, their study with the most powerful telescopes can map the universe back to the very first stars, believed to be very massive. GRBs from such large redshifts will require a next generation telescope to follow the *Swift* mission, currently planned for launch in 2003. The proposed Energetic X-ray Imaging Survey Telescope, *EXIST*, can achieve these objectives as the Next Generation GRB mission.

We describe the current mission concept for *EXIST* and then some of the GRB and associated science that could be conducted with a nominal 5y mission.

## *EXIST* MISSION CONCEPT

The primary mission requirements for a next generation GRB mission are: 1. very large area, for maximum GRB sensitivity; 2. large field of view (FOV), for collection of a large GRB sample and sensitivity to rare events, including low-luminosity nearby GRBs possibly observed off-axis (e.g. SN1998bw); 3. high angular resolution and fine positional determination (10") for (near-) real time optical identifications; and 4. broad energy band coverage, with imaging response down to ~10 keV for high z GRBs and up to ~600 keV to extend beyond the ~300 keV $\nu F\nu$ energy peak and maximize sensitivity to broad energy-dependent lags (hard-soft) which may allow measures of GRB luminosity-distance.

Broad-band (~10-600 keV) hard x-ray imaging over a wide FOV is best conducted with a coded aperture telescope with an imaging detector capable of fine position resolution (to record the coded mask shadow), high Z stopping power for good high energy sensitivity in moderate detector thickness, and compact mounting and tiling capability for extension to a very large area total detector array. Coded aperture imagers are background limited and so record signal to noise S/N ~ $(A_{det} \cdot T/B)^{0.5}$, for a given detector area $A_{det}$ recording background B (cts cm$^{-2}$sec$^{-1}$) for a source observed for time T(sec). Optimum imaging sensitivity requires systematic variations on the detector (e.g., due to gain and non-uniform background variations) to be effectively averaged, which can best be achieved by continuously scanning the detector-telescope across the sky. Since for the wide-field (>>10°) imaging needed for GRB sample statistics, the background is dominated by the diffuse cosmic flux (primarily over ~20-200 keV) recorded in the FOV of size θ x θ, then B~ $\theta^2$. Similarly, the exposed $A_{det}$, available integration time T, and recorded B are each proportional to the angular width θ of the FOV in the scan direction. Thus, for a scan at orbital rate dφ/dt ~4° min$^{-1}$, the expected S/N ~ $\theta^{0.5}$ for GRBs with duration $T_b$ ~ θ(dφ/dt)$^{-1}$ ~ 3-10min (e.g. long GRBs, or possibly high-z GRBs) and S/N ~ independent of θ for typical GRBs ($T_b$ < 1min). Since the total GRB sample $N_b \sim \theta^2$, large $\theta$ is optimum; this also maximizes the persistent source sensitivity.

These general considerations, as well as the primary goal to conduct a hard x-ray imaging survey which extends ROSAT sensitivity to >100keV but with all sky coverage each orbit [1], yield the preliminary *EXIST* mission concept outlined below. The baseline implementation is for a Free Flyer mission although a version studied originally [1,2] could be mounted on the International Space Station.

## Mission Implementation Overview

Three telescopes, each with a 60º x 75º fully-coded FOV, are mounted on a base spacecraft as shown in Figure 1a to form a combined 180º x 75º fan beam which images the full sky each orbit. Each telescope is constructed (Figure 1b) of a 3 x 3 array of actively collimated (CsI) overlapping FOV (60º x 50º) sub-telescope modules, each read out with arrays of Cd-Zn-Te (CZT) detectors (≥5mm thick; 1.3mm pixels) viewing the sky through canted-flat coded apertures.

**Table 1:** *EXIST* **Mission Parameters**

| | |
|---|---|
| Energy range | 10-600 keV |
| FOV | 180º x 75º (fully coded) |
| | ~5 steradians (partial coded) |
| Angular Resolution | 2-5' (10-50"source locations) |
| Energy/Temporal Resolution | 1-3%; 2 μsec |
| Sensitivity (5σ, ≤1y) | ~0.05 mCrab (10-100 keV) |
| | ~0.5 mCrab (>200 keV) |
| Telescopes, Detectors | Coded aperture, 8 m² CZT |
| Pointing, Aspect | ~1° stability, 5" knowledge |
| Mass, Power, TM | 8500kg, 1500W, 1.5Mbs |
| Launch, Cost (incl. Ops) | Delta IV, $330M (w/ cont.) |

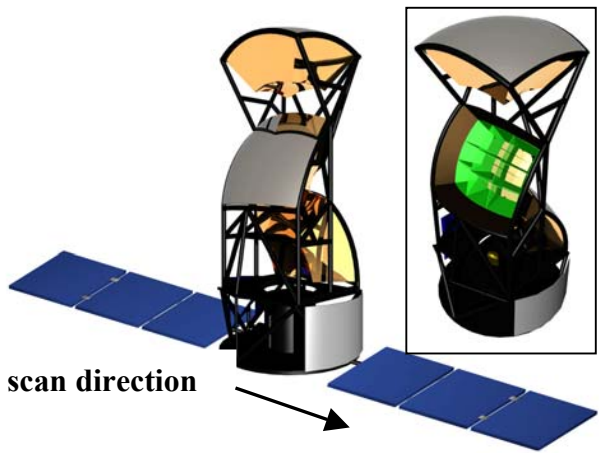
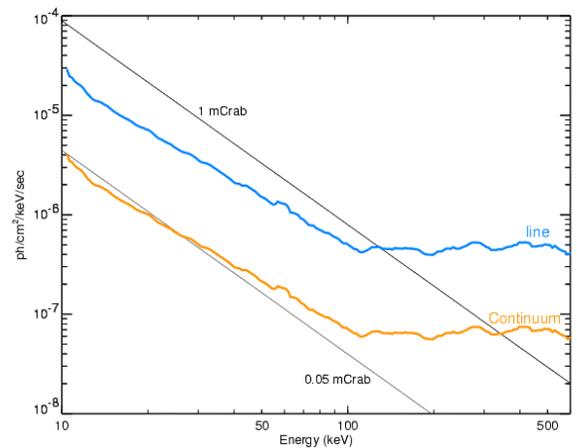

**FIGURE 1. a)** *EXIST* telescopes (3) on spacecraft, zenith pointed along orbital scan direction, **b)** cutaway view showing imaging CZT detector arrays and active collimator.

**FIGURE 2.** *EXIST* survey sensitivities (5σ). GRB sensitivity is ~50mCrab for an assumed 10sec duration burst.

Parameters for the mission are summarized in Table 1 and estimated continuum and line sensitivities (5σ, ≤1y, depending on source orbital latitude) are shown in Figure 2. After the first year all-sky survey, or at any time for Targets of Opportunity, the mission would be operated as an Observatory, with an active guest observer program for pointed observations with the central telescope while the outer two telescopes continue the Survey. GRBs are imaged in the wide combined field (5sr) regardless of Survey or Observatory mode. The active collimator and rear shield (1cm and 2cm thick CsI, respectively) will be pulse-height-analyzed to extend GRB spectral coverage up to ~1-3MeV. Burst positions are derived to ~1-3' within 10sec for rapid transmission to ground and other observatories, and to within ~10-50" in ground analysis of the full data within ~3hours.

Primary Survey science is the study of obscured AGN and black holes on all scales. Details of the full mission and science will be given in a later paper.

# GRB SCIENCE FROM EXIST

*EXIST* is a Next Generation Burst Observatory. With sensitivity to weak events a factor ~20 below BATSE and ~3 – 10 below *Swift* (given the extended low and high energy band of *EXIST*), it should provide ~10 – 50'' locations for 2 – 3 GRBs a day. With its large instantaneous field of view it can study both low luminosity nearby GRB events (like SN1998bw) as well as the brightest events most likely to be observed by gravitational wave and neutrino detectors. *EXIST* will be on orbit at an amazing time, when *NGST* will be studying the high-redshift universe and *Contellation-X* will enable high resolution spectra of x-ray afterglows pinpointed by *EXIST* to be measured. *LISA* and LIGO2 will be providing the first sensitive gravitational wave detections, and IceCube, Auger, and other high-energy neutrino and ultra-high energy cosmic ray detectors will be operating. The coincidence of these capabilities with *EXIST* will provide opportunities to search for the first massive stars in the universe to very high redshifts and open up non-electromagnetic channels of GRB energy release for observation – challenging our theories of relativistic shocks and testing physics from special relativity to neutrino masses and couplings.

## First Massive Stars to *EXIST*

Nearly 2/3 of all GRBs, the "long bursts", explode at significant cosmological distances. With its high sensitivity, EXIST can detect GRBs at high redshift (z~10-20; cf. Figure 3), enabling the first direct search for the initial generation of stars (Pop III) that were likely very massive. Their resulting epoch of black hole (BH) formation would, given increasing evidence for long burst (>2sec) GRB production by hypernovae in Collapsars [4, 5], produce an epoch of GRBs. Spectroscopy of their optical afterglows, detectable in the IR with NGST, would enable mapping cosmic structure back to the "dark ages" [6]. This requires a rapid estimate of GRB redshift, which could be provided by the "photometric redshifts" to be derived from the observed relation [7] (cf. Figure 4) between GRB hard-soft lags and absolute luminosity.

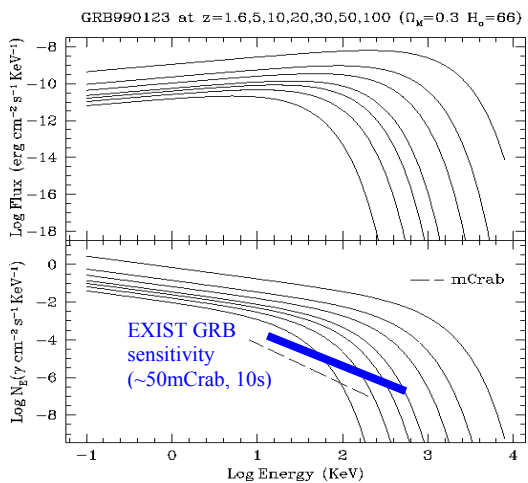

**FIGURE 3.** GRB sensitivity vs. z (adapted from [3]).

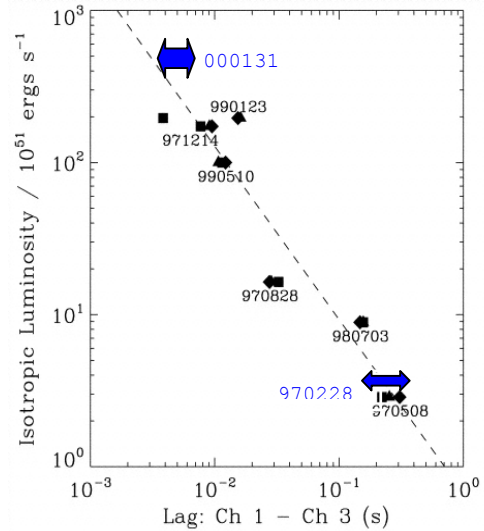

**FIGURE 4.** Luminosity-lag relation for GRBs (adapted from [7]). The observed correlation, $L_{pk} \propto \Delta\tau^{-1.14}$, may be understood as a kinematic effect from beaming [8].

Application of this technique to GRBs at high z requires the very large collection area for optimum statistics as well as the broad energy band coverage and good spectral resolution of *EXIST*.

The large GRB sample with approximate redshifts collected by *EXIST* will constrain the star formation rate to large z, complementing *NGST*. The *EXIST* deep sample will also provide possible GRB locations and redshifts for "orphan afterglows", probably found with SDSS [9] and expected in quantity from LSST in the *EXIST* era. Since orphans may preferentially be off-axis GRBs [10], they are likely to be x-ray bright and thus possibly related to the x-ray flashes (XRFs) which appear to form a spectral extension to GRBs [11]. Alternatively, the XRFs may be GRBs from massive (~300$M_\odot$) Pop III stars at z ~ 10 undergoing collapse to BHs, which also would produce background high energy (~$10^5$ GeV) neutrinos detectable with IceCube [12].

## Highest Energy Particles to *EXIST*

The relativistic shocks driven into the interstellar medium after a GRB explosion can produce high-energy neutrinos in several components with energies ~10 GeV, ~100 TeV and ~$10^{18}$ eV on timescales of <10s. Theory [13] predicts that several tens of muon induced neutrino events/year should be detectable with the IceCube neutrino telescope in coincidence with GRBs. Detection of the high energy neutrinos would not only test the shock acceleration mechanism, but would also imply GRBs are sources of Ultra High Energy Cosmic Rays (UHECR). A GRB at 100 Mpc (for which the wide-field, high sensitivity *EXIST* trigger is needed) producing 100 TeV neutrinos also allows a test of neutrino mass mixing five orders of magnitude more sensitive than solar neutrinos.

## Short GRBs and SGRs

Short (<2sec) GRBs are still unidentified with afterglows or hosts and are thus without luminosities or redshifts. They are consistent with arising in NS-NS (or perhaps NS-BH) mergers [5]. *EXIST* should detect ~$10^3$ over a 3 year mission with fluxes above the BATSE threshold and locate these to ~10" positional accuracy, sufficient to determine whether they are also (like long GRBs) associated with star formation regions in galaxies. If they arise from NS-NS mergers, or from accretion induced collapse of WDs [14], they may preferentially arise in globular cluster systems. Studies of the light curve variability enabled by the high statistics of *EXIST* can constrain short GRB jets (testing AIC), as it has apparently for long GRBs [15].

The soft gamma-ray repeaters (SGRs) are detected in the Galaxy (3) and LMC (1) and are associated with NSs, probably magnetars [16]. The Giant Flares detected from two of these reach highly super-Eddington luminosities which *EXIST* could detect with its all-sky monitoring and imaging out to ~3-10Mpc. This would survey SGRs and the incidence and activity of magnetars throughout the Local Group.

## CONCLUSIONS

A Next Generation GRB mission to conduct the highest sensitivity direct study of GRBs would culminate the steady advance in sensitivity, resolution (spatial and spectral) and broad sky coverage, from BATSE through HETE-2 and *Swift*. The proposed *EXIST* mission would achieve this. It would provide the high sensitivity-resolution spectra and images for high energy GRBs from *GLAST* and afterglows from *Constellation-X*, *NGST*, *LSST*, and very large ground-based telescopes. As recommended by the Decadal Survey, *EXIST* could be launched by 2010 to open a new frontier in the study of the most extreme cosmic phenomena and the first stellar objects.